\documentclass[twocolumn,prd,amsmath,superscriptaddress]{revtex4}
\usepackage{graphicx}
\usepackage{latexsym}
\usepackage{amsmath}
\usepackage{amssymb}
\usepackage{graphics}
\usepackage{color}
\usepackage{ifthen}
\usepackage{bm}
\usepackage{color}

\definecolor{darkgreen}{RGB}{0,180,0}

\definecolor{darkblue}{rgb}{0.20,0.3,0.6}
\usepackage[linktocpage=true]{hyperref}
\hypersetup{
colorlinks=true,
citecolor=darkblue,
linkcolor=darkblue,
urlcolor=darkblue,
pdfauthor={},
pdftitle={},
pdfsubject={}
}

\newcommand\ee{\end{equation}}
\newcommand\be{\begin{equation}}
\newcommand\eea{\end{eqnarray}}
\newcommand\bea{\begin{eqnarray}}

\def\gsim{ \lower .75ex \hbox{$\sim$} \llap{\raise .27ex \hbox{$>$}} }
\def\lsim{ \lower .75ex \hbox{$\sim$} \llap{\raise .27ex \hbox{$<$}} }

\begin{document}

\title{DBI Genesis: An Improved Violation of the Null Energy Condition}
\author{Kurt Hinterbichler}
\affiliation{Perimeter Institute for Theoretical Physics, 31 Caroline St. N, Waterloo, Ontario, Canada, N2L 2Y5}

\author{Austin Joyce}
\author{Justin Khoury}
\author{Godfrey E. J. Miller}
\affiliation{Center for Particle Cosmology, University of Pennsylvania, Philadelphia, PA 19104 }%


\begin{abstract}
We show that the DBI conformal galileons, derived from the world-volume theory of a 3-brane moving in an AdS
bulk, admit a background, stable under quantum corrections, which violates the Null Energy Condition (NEC).  The perturbations around this background are stable and
propagate subluminally.  Unlike other known examples of NEC violation, such as ghost condensation and conformal galileons, this theory also admits
a stable, Poincar\'e-invariant vacuum, with a Lorentz-invariant S-matrix satisfying standard analyticity conditions. Like conformal galileons, perturbations
around deformations of the Poincar\'e invariant vacuum propagate superluminally.
\end{abstract}

\maketitle

The NEC is the most robust of all energy conditions. It states that, for any null vector $n^\mu$,
\be
T_{\mu\nu} n^\mu n^\nu \geq 0\,.
\label{nec}
\ee
It has proven extremely difficult to violate this condition with well-behaved relativistic quantum field theories.
Aside from being of purely theoretical interest, the NEC
plays a role in our understanding of the early universe.  In cosmology,~\eqref{nec} is equivalent to
$\rho + P \geq 0$, which, combined with the equation for a spatially-flat universe,
\be
M_{\rm Pl}^2 \dot{H} =  -\frac{1}{2}(\rho + P)\,,
\label{dotHeq}
\ee
forbids a non-singular bounce from contraction to expansion.  This means a contracting universe necessarily ends in a big crunch singularity, and
an expanding universe must emerge from a big bang. Violating~(\ref{nec}) is therefore central to any alternative to inflation relying either
on a contracting phase before the big bang~\cite{Gasperini:1992em,Khoury:2001wf,Rubakov:2009np,Hinterbichler:2011qk,Hinterbichler:2012mv}, or an expanding phase from an asymptotically static past~\cite{Nayeri:2005ck,Creminelli:2010ba}.

For theories with at most two derivatives, violating the NEC necessarily implies ghosts or gradient instabilities~\cite{Dubovsky:2005xd}. To evade this, one must therefore invoke higher derivatives, as in the {\it ghost condensate}~\cite{ArkaniHamed:2003uy}. Perturbations around the ghost condensate can violate the NEC in a stable manner~\cite{Creminelli:2006xe},
and this has been used in the New Ekpyrotic scenario~\cite{Buchbinder:2007ad,Creminelli:2007aq}.
However, because the scalar field starts out with a wrong-sign kinetic term, the theory is unstable around its Poincar\'e-invariant vacuum.

Stable NEC violation can also be achieved with {\it conformal galileons}~\cite{Nicolis:2008in}, a class of conformally-invariant scalar field theories with particular
higher-derivative interactions. Remarkably, in spite of the fact that there are five independent galileon terms, only the kinetic term contributes to~\eqref{nec} \cite{Creminelli:2012my}:
violating the NEC requires a wrong-sign kinetic term, just like the ghost condensate. Another issue with conformal galileons is superluminal propagation around slight deformations
of the NEC-violating background \cite{Creminelli:2010ba} (though this can be avoided by explicitly breaking special conformal transformations~\cite{Creminelli:2012my}).

In this Letter, we show that the {\it DBI conformal galileons}~\cite{deRham:2010eu,Goon:2011qf} can also violate the NEC in a stable manner, while avoiding nearly all of the aforementioned issues. Specifically, the coefficients of the five
DBI galileons can be chosen such that:
\begin{enumerate}

\item There exists a stable, Poincar\'e-invariant vacuum. 

\item The Lorentz-invariant S-matrix about this vacuum obeys standard analyticity conditions.

\item The theory admits a time-dependent, homogeneous and isotropic solution which violates the NEC in a stable manner.

\item Perturbations around the NEC-violating background, and around small deformations thereof, propagate subluminally.

\item This solution is stable against radiative corrections.

\end{enumerate}

In other words, starting from a local relativistic quantum field theory defined around a Poincar\'e-invariant vacuum state, the theory allows consistent, stable, NEC-violating solutions. 
In fact, this NEC-violating background is an {\it exact} solution of the effective theory, including all possible higher-dimensional operators consistent with the assumed symmetries. 

We will see that the above conditions can be satisfied for a broad region of parameter space. This represents a significant improvement over ghost condensation (which fails to satisfy 1 and 2) and the ordinary conformal galileons (which fail to satisfy 1, 2  and 4). Unfortunately, like conformal galileons, superluminal propagation around deformations of the Poincar\'e invariant solution is inevitable. 
Additionally, one would like the theory to be consistent with black hole thermodynamics~\cite{Dubovsky:2006vk}. This is currently under investigation~\cite{yizen}. 

The geometric origin of the DBI conformal galileon as the theory of a 3-brane moving in
an AdS$_5$ bulk makes contact with stringy scenarios, offering a promising avenue to search for NEC violations in string theory.

{\it The theory}: Consider a 3-brane, with worldvolume coordinates $x^\mu$, probing an AdS$_5$ space-time
with coordinates $X^A$ and metric $G_{AB}(X)$ in the Poincar\'e patch 
\be
{\rm d}s^2 = G_{A B}{\rm d}X^A{\rm d}X^B= Z^{-2}{\rm d}Z^2+Z^2\eta_{\mu\nu}{\rm d}X^\mu {\rm d}X^\nu  \,,
\ee 
where $Z\equiv X^5$, $0<Z< \infty$. The dynamical variables are the embedding functions, $X^\mu(x)$, $Z(x)\equiv \phi(x)$.
In unitary gauge, $X^\mu = x^\mu$, the brane induced metric is
\be
\mathfrak{g}_{\mu\nu} = G_{AB} \partial_\mu X^A \partial_\nu X^B  = \phi^2 \eta_{\mu \nu} + \phi^{-2} \partial_\mu \phi\partial_\nu \phi \,.
\ee
The DBI conformal galileon action is a sum of five geometric invariants, with free coefficients $c_1,\ldots, c_5$:
\bea
\nonumber
{\cal L} &=& c_1{\cal L}_1  + c_2 {\cal L}_2 + c_3 {\cal L}_3 + c_4 {\cal L}_4 + c_5 {\cal L}_5\,, ~~~{\rm where}\\
\nonumber
{\cal L}_1 &=& -\frac{1}{4} \phi^4\,,  \\
\nonumber
{\cal L}_2 &=& -\sqrt{-\mathfrak{g}} = - \gamma^{-1} \phi^4\,, \\
\nonumber
{\cal L}_3 &=& \sqrt{-\mathfrak{g}} K = - 6 \phi^4 + \phi [\Phi]
+ \gamma^2 \phi^{-3} \left( - [\phi^3] + 2 \phi^7 \right) ,\\
\nonumber
{\cal L}_4 &=& - \sqrt{-\mathfrak{g}} \mathfrak{R} \\
\nonumber
&=& 12 \gamma^{-1} \phi^4 + \gamma \phi^{-2} \left\{ [\Phi^2] - \left( [\Phi] - 6 \phi^3 \right) \left( [\Phi] - 4 \phi^3 \right) \right\}\\
\nonumber
&+&  2 \gamma^3 \phi^{-6} \left\{ - [\phi^4] + [\phi^3] \left( [\Phi] - 5 \phi^3 \right) - 2 [\Phi] \phi^7 + 6 \phi^{10} \right\} ,\\
\nonumber
{\cal L}_5 &=& \frac{3}{2} \sqrt{-\mathfrak{g}}   \left(- \frac{K^3}{3}  + K_{\mu \nu}^2 K - \frac{2}{3} K_{\mu \nu}^3 - 2 \mathfrak{G}_{\mu \nu} K^{\mu \nu}\right) \\
\nonumber
&=& 54 \phi^4 - 9 \phi [\Phi]  + \gamma^2 \phi^{-5} \left\{ 9 [\phi^3] \phi^2 + 2 [\Phi^3] - 3 [\Phi^2] [\Phi] \right. \\
\nonumber
&+ & \left. 12 [\Phi^2] \phi^3 + [\Phi]^3 - 12 [\Phi]^2 \phi^3 + 42 [\Phi] \phi^6 - 78 \phi^4 \right\}  \\
\nonumber
&+&  3 \gamma^4 \phi^{-9} \left\{ -2 [\phi^5] + 2 [\phi^4] \left( [\Phi] - 4 \phi^3 \right) \right. \\
\nonumber
&+& [\phi^3] \left( [\Phi^2] - [\Phi]^2 + 8 [\Phi] \phi^3 - 14 \phi^6 \right) \\
&+& \left. 2 \phi^7 \left( [\Phi]^2 - [\Phi^2] \right) - 8 [\Phi] \phi^{10} + 12 \phi^{13} \right\}
.
\label{Ls}
\eea
Here $\gamma\equiv 1/\sqrt{1 + (\partial\phi)^2/\phi^4}$ is the Lorentz factor for the brane motion, ${\cal L}_1$ measures the proper 5-volume between the brane and some fixed
reference brane~\cite{Goon:2011qf}, and ${\cal L}_2$ is the world-volume action, {\it i.e.}, the brane tension~\cite{footnote1}. 
The higher-order terms ${\cal L}_3$, ${\cal L}_4$ and ${\cal L}_5$ are functions of the extrinsic curvature tensor $K_{\mu\nu} = \gamma\left(-\phi^{-1}\partial_\mu\partial_\nu\phi  + \phi^2\eta_{\mu\nu} + 3\phi^{-2} \partial_\mu\phi\partial_\nu\phi\right)$ and the induced Ricci tensor $\mathfrak{R}_{\mu\nu}$ and scalar $\mathfrak{R}$, with $\mathfrak{G}_{\mu \nu} \equiv \mathfrak{R}_{\mu\nu} - \mathfrak{R} \mathfrak{g}_{\mu\nu}/2$ (and indices raised by $\mathfrak{g}^{\mu\nu}$).  Following~\cite{Goon:2011qf}, $\Phi$ denotes the matrix of second derivatives $\partial_\mu\partial_\nu \phi$, $[\Phi^n] \equiv {\rm Tr}(\Phi^n)$, and $[\phi^n] \equiv \partial\phi \cdot \Phi^{n-2} \cdot \partial\phi$, with indices raised by $\eta^{\mu\nu}$.

Each ${\cal L}$ is invariant up to a total derivative under the $so(4,2)$ conformal algebra, inherited from the isometries of AdS$_5$.
Aside from Poincar\'e transformations,~\eqref{Ls} is also invariant under dilation, $\delta_D\phi = -(1 + x^\mu\partial_\mu)\phi$, and special conformal transformations,
$\delta_{K_\mu} \phi = (-2x_\mu - 2x_\mu x^\nu\partial_\nu + x^2\partial_\mu + \phi^{-2}\partial_\mu )\phi$.

{\it Around the Poincar\'e Invariant Vacuum}: Expanding~\eqref{Ls} around a constant field profile, $\bar{\phi}_0$, up to quartic order in
perturbations $\varphi = \phi - \bar{\phi}_0$, we obtain
\bea
\nonumber
{\cal L}&=& - \frac{C_2}{2}(\partial\varphi)^2 + \frac{C_3 }{12 \bar{\phi}_0^3} (\partial \varphi)^2 \square \varphi + \frac{( 3 C_2 - C_3 )}{24\bar{\phi}_0^4} (\partial \varphi)^4 \\
\nonumber
&  - &  \frac{C_3}{4\bar{\phi}_0^4} \varphi (\partial \varphi)^2 \square \varphi \ +  \frac{C_4}{24 \bar{\phi}_0^6} (\partial \varphi)^2 \left[(\partial_\mu\partial_\nu \varphi)^2 - (\square \varphi)^2\right]\,; \\
\nonumber
&C_2& \equiv  c_2 + 6 c_3 + 12 c_4 + 6 c_5\;, ~~ C_3 \equiv  6 c_3 + 36 c_4 + 54 c_5\;, \\
&C_4& \equiv  12 c_4 + 48 c_5\;,\;\;\;\;\;\;\;\;\;\;\;\;\;\;\;\;\;\, C_5 \equiv c_5 \,,
\label{Poincaretheory}
\eea
%
where, in order for $\bar{\phi}_0$ to be a solution, we have imposed that the tadpole term vanish:
\be
C_1 \equiv - \frac{1}{4} c_1 - c_2 - 4 c_3 + 12 c_5 = 0~~({\rm Poincar\acute{e}}~{\rm solution})\,.
\label{C1cond}
\ee
Stability of small fluctuations requires
\be
C_2 > 0 \qquad ({\rm stability})\,.
\label{C2cond}
\ee
%

Next, the scattering S-matrix derived from~\eqref{Poincaretheory} should satisfy standard relativistic dispersion relations. Firstly, the $2\to 2$ amplitude in the forward limit must display a positive $s^2$ contribution~\cite{nimaUVIR}. Only the $(\partial\varphi)^4$ vertex contributes in the forward limit --- its coefficient must be strictly positive~\cite{nimaUVIR,Komargodski:2011vj}. There also exist constraints away from the forward limit~\cite{Martin:1965jj}, which involve the $(\partial \varphi)^2 \square \varphi$ and $(\partial \varphi)^2(\partial_\mu\partial_\nu \varphi)^2$ vertices~\cite{Nicolis:2009qm}. These analyticity conditions respectively impose
\be
C_3 < 3C_2\;; \qquad C_3^2 > 6C_2C_4 \qquad ({\rm analyticity})\,.
\label{C3analyticity}
\ee
%

{\it NEC-Violating Solution}: We seek a time-dependent, isotropic background solution of the form
\be
\bar{\phi} = \frac{\alpha}{(-t)}\,;\qquad -\infty < t < 0\,,
\label{phibar}
\ee
where $\alpha$ is a constant. This profile, which is central to pseudo-conformal~\cite{Rubakov:2009np,Hinterbichler:2011qk,Hinterbichler:2012fr} and Galilean Genesis~\cite{Creminelli:2010ba} cosmology,
spontaneously breaks the $so(4,2)$ algebra down to an $so(4,1)$ subalgebra. Substituting~(\ref{phibar}) into the equation of motion for $\phi$ derived from~\eqref{Ls}, we obtain
\be
C_2 + \frac{1}{2}C_3\beta + \frac{1}{2} C_4 \beta^2 + 6 C_5 \beta^3= 0 \qquad (1/t~{\rm solution}) \,,
\label{alphaeom}
\ee
with $\beta \equiv  \bar{\gamma} - 1 > 0$, $\bar{\gamma} = 1/\sqrt{1-\alpha^{-2}}$.
There is a solution for each real, positive root of~\eqref{alphaeom}.

We require this background to be stable against small perturbations. 
Expanding~\eqref{Ls} to quadratic order in $\varphi \equiv \phi - \bar{\phi}$, we obtain
\be
{\cal L}_{{\rm quad},~1/t} = \frac{{\cal Z}}{2}  \left( \dot{\varphi}^2 - \bar{\gamma}^{-2}(\vec{\nabla} \varphi)^2 + \frac{6}{t^2} \varphi^2\right) \,,
\label{Lquad}
\ee
where ${\cal Z} \equiv  \bar{\gamma}^3 (C_2 + C_3 \beta + 3C_4 \beta^2/2 + 24 C_5 \beta^3)$.
Absence of ghosts therefore requires
\be
C_2 + C_3 \beta + \frac{3}{2} C_4 \beta^2 + 24 C_5 \beta^3 > 0 \qquad ({\rm stability})\,.
\label{stabtimedep}
\ee
The sound speed is always subluminal, but for small deformations away from the solution to satisfy Condition~4, we want the sound speed $c_s =  \bar{\gamma}^{-1}$ to be generously less than unity.  Thus we demand
\be
\beta \; \gsim \; 1\qquad ({\rm robust\ subluminality}~{\rm around}~1/t)\,.
\label{sublum}
\ee

To check for NEC violation, we calculate the stress tensor $T_{\mu\nu}$ by varying the covariant version of~\eqref{Ls} with respect to the metric. The covariant theory is given uniquely by the brane construction~\cite{Goon:2011qf}, and is given by~\eqref{Ls} with the replacements $\eta_{\mu\nu}\rightarrow g_{\mu\nu}$ and $\partial_\mu\rightarrow \nabla_\mu$, plus the following non-linear couplings:
\bea
\nonumber
\delta {\cal L}_4 &=& - \gamma^{-1} R \phi^2 + 2 \gamma \phi^{-2} R^{\mu \nu} \nabla_\mu \phi \nabla_\nu \phi \\
\nonumber
\delta {\cal L}_5 &=& (3/2) R \phi^{-5} \left\{ \phi^4 \left( [\Phi] - 4 \phi^3 \right) + \gamma^2 \left( - [\phi^3] + 2 \phi^7 \right) \right\}   \\ 
\nonumber
&-& 3 \phi^{-1} R^{\mu\nu} \nabla_\mu \nabla_\nu \phi
\\
\nonumber
&+&  3 \gamma^2 \phi^{-5} R^{\mu \nu}  \left( \left( 4 \phi^3 - [\Phi] \right) \nabla_\mu \phi + \nabla^\kappa \phi \nabla_\kappa \nabla_\mu \phi \right) \nabla_\nu \phi  \\
& + & 3 \gamma^2 \phi^{-5} R^{\mu \kappa \nu \lambda} \nabla_\mu \phi \nabla_\nu \phi \nabla_\kappa \nabla_\lambda \phi \,,
\label{delLs}
\eea
where indices are now raised and lowered with $g_{\mu\nu}$, and we assume an overall $\sqrt{-g}$ factor. Varying the action with respect to the metric, and evaluating the result on the solution
$\bar{g}_{\mu\nu}  = \eta_{\mu\nu}$ and $\bar{\phi} = \alpha/(-t)$, yields an isotropic $T_{\mu\nu}$, with vanishing energy density and pressure scaling as $t^{-4}$ (as it must by dilation invariance~\cite{Creminelli:2010ba,Hinterbichler:2012mv}),
\be
\rho = 0\,;\ \ \ P  = \frac{\alpha^2}{t^4} \left(C_2 - C_4 + 12 C_5 \right)\,,
\ee
where we have used~(\ref{alphaeom}) to simplify. To violate the NEC, the pressure must be negative,
\be
C_2 - C_4 + 12 C_5 < 0\qquad ({\rm NEC}\ {\rm violation})\,.
\label{NECviolcond}
\ee

{\it Matching to Standard Cosmology}: By coupling this sector minimally to Einstein-Hilbert gravity, we obtain a {\it DBI Genesis} cosmology. Integrating~\eqref{dotHeq} yields
\be
H(t) = - (C_2 - C_4 + 12 C_5) \frac{\alpha^2}{3M_{\rm Pl}^2(-t)^3}\,,
\label{HGen}
\ee
which describes an expanding universe from an asymptotically static state. 

For this to represent a useful violation of the NEC, we must verify that the DBI Genesis phase can match onto a standard, {\it expanding} radiation-dominated phase.
We remain agnostic about the details of the reheating process; our main concern is whether the
universe is expanding after the transition. While one might expect that $H$ matches continuously if the transition is fast enough,
this is not so~\cite{Creminelli:2012my} --- the pressure includes a contribution, $P_{\rm sing} \sim \ddot{\phi}$, which diverges as $\phi$ is brought
instantaneously to a halt. In our case, we obtain
\begin{align}
\nonumber
&~~~~~~~~~~~~~~~~~~~~~~P_{\rm sing}  = \dot{F}~;~~~~{\rm where}\\\nonumber
&F(t) \equiv  \frac{\alpha^2}{6 (-t)^3} \bigg(24 C_5- 2 C_4 - (2 C_4 - 60 C_5) \beta-18 C_5\beta^2   \\\
& ~~~~~-( C_3 - 3 C_4 + 90 C_5)\left( \frac{\bar{\gamma}\cosh^{-1} \bar{\gamma} }{\sqrt{1+\bar{\gamma}} \sqrt{\beta}} - 1 \right) \bigg)\,.
\label{Fdef}
\end{align}
%
The conserved quantity is not $H$, but rather $H + F/2M_{\rm Pl}^2$. In other words, neglecting the $\phi$ contribution in the
post-Genesis universe, the matching condition at reheating is
\be
H_{\rm Genesis} + \frac{F}{2M_{\rm Pl}^2} = H_{\rm rad.-dom.} \,.
\ee
Combining~(\ref{HGen}) and~(\ref{Fdef}), we find that the universe will be expanding in the radiation-dominated phase if
\bea
\nonumber
& & 2C_2 + ( 2 C_4 - 60 C_5 ) \beta + 18 C_5 \beta^2  +  ( C_3 - 3 C_4 + 90 C_5 )\\
& & \times \left( \frac{\bar{\gamma}\cosh^{-1} \bar{\gamma} }{\sqrt{1+\bar{\gamma}} \sqrt{\beta}} - 1 \right)  < 0~~~~~~({\rm matching}) \,.
\label{matchcond}
\eea

{\it Summary of Conditions}: We started out with five coefficients, $C_1,\ldots, C_5$. Stability of the Poincar\'e-invariant vacuum sets $C_1 = 0$
and (without loss of generality) $C_2 = 1$. This leaves us with three coefficients, $C_3$, $C_4$ and $C_5$, which must be chosen such
that the cubic equation~\eqref{alphaeom} has a real root with $\beta\;\gsim\; 1$ (per~(\ref{sublum})),
and which must satisfy the inequalities~\eqref{C3analyticity},~\eqref{stabtimedep},~\eqref{NECviolcond} and~\eqref{matchcond}.

\begin{figure} 
   \centering
   \includegraphics[width=0.3\textwidth]{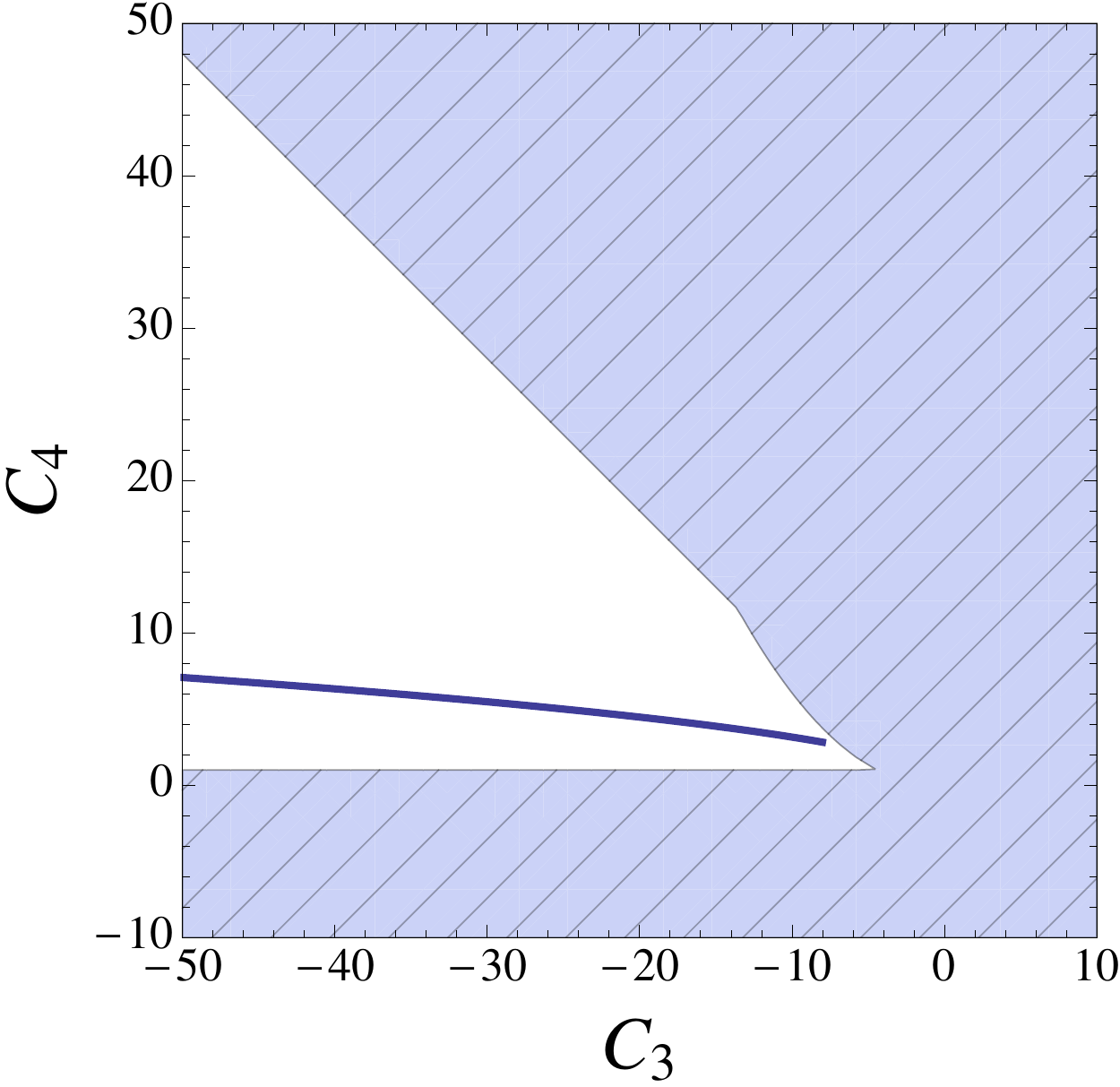}
   \caption{Allowed (white) region of $(C_3,C_4)$ parameter space satisfying all of our conditions, with $C_1 = C_5 = 0$ and $C_2 = 1$.
   In the allowed region, $\beta \simeq -C_3/C_4$ for $|C_3| \gg 1$.  On the solid curve, $\beta$ grows without bound as $C_3 \rightarrow -\infty$, showing that all constraints can be satisfied for arbitrarily large $\beta$.}
   \label{C3C4plot}
\end{figure}

All these conditions can be satisfied even with $C_5 = 0$. With $C_2 = 1$, the first inequality in~\eqref{C3analyticity} gives $C_3 < 3$, while~\eqref{NECviolcond} simplifies to
$C_4 > 1$. The equation of motion~\eqref{alphaeom} reduces to a quadratic equation, with roots $\beta_\pm = (\pm \sqrt{C_3^2 - 8 C_4} - C_3)/2 C_4$. It is easy to check that
only $\beta_+$ can lead to a stable $1/t$ solution. In order for $\beta_+$ to be real and $\gsim\; 1$, we must require $C_3^2 > 8C_4$ and $C_3 \; \lsim -(2+C_4)$. 
With these conditions,~\eqref{stabtimedep} and the second inequality of~\eqref{C3analyticity} are automatically satisfied. The only remaining constraint is~\eqref{matchcond}. Figure~\ref{C3C4plot} shows (in white) the allowed region of $(C_3,C_4)$ parameter space satisfying all of our constraints.
Generalizing the analysis to $C_5 \neq 0$ only widens the allowed region.

{\it Quantum Stability}. We now argue that the NEC-violating solution is robust against other allowed terms in the
effective theory, {\it i.e.}, all diffeomorphism invariants of the induced metric and extrinsic curvature. Using the Gauss-Codazzi relation $\mathfrak{R}_{\mu\nu \rho\sigma} = \frac{2}{3}(\mathfrak{g}_{\mu\rho}\mathfrak{g}_{\nu\sigma} - \mathfrak{g}_{\mu\sigma}\mathfrak{g}_{\nu\rho}) +  K_{\mu\rho} K_{\nu\sigma} - K_{\mu\sigma}K_{\nu\rho}$ to eliminate all instances of $\mathfrak{R}_{\mu\nu\rho\sigma}$ in favor of $K_{\mu\nu}$, we see that the DBI galileons are particular polynomials in $K_{\mu\nu}$. 
As argued in the Appendix of~\cite{Khoury:2012dn}, however, any polynomial in $K_{\mu\nu}$ can be brought to the galileon form through field redefinitions. 

It remains to consider terms with covariant derivatives acting on $K_{\mu\nu}$, such as $K_{\mu\nu}\square K^{\mu\nu}$.
Since $\bar{K}_{\mu\nu} = - \bar{\gamma} \bar{\mathfrak{g}}_{\mu\nu}$ on the $1/t$ background, it is annihilated by $\nabla$, so
these higher-derivative terms do not contribute to the equation of motion for the $1/t$ ansatz. Hence the $1/t$ solution is an {\it exact} solution,
including all possible higher-derivative terms in the effective theory.

These higher-derivative terms {\it do} contribute to perturbations, but it is technically natural to set their coefficients to zero if there is a hierarchy, $C_3 \sim \beta$,  $C_2 \sim C_4 \sim {\cal O}(1)$, $C_5 \sim 1/\beta$,
where $\beta \gg 1$ ($\alpha \simeq 1$). This corresponds to relativistic brane motion. The solid curve in Fig.~\ref{C3C4plot}, corresponding to $C_4 \simeq -C_3/\beta$ for $\beta \gg 1$, shows that all of our constraints can be satisfied for arbitrarily large $\beta$. In the limit of large $|t|$, the theory of perturbations is approximately the same as that about a constant background. Consequently, the fluctuation lagrangian takes the form~\eqref{Poincaretheory}, where now $\bar\phi_0$ is~\eqref{phibar}, except that every spatial gradient is multiplied by a factor of the sound speed, $1/\bar{\gamma}\simeq 1/\beta$. A computation shows that the coefficient of an ${\cal O}(\varphi^n)$ term scales as $\beta^{2n+1}$. The (ordinary) galileon terms are suppressed by the lowest scale in the theory
\be
\Lambda_{\rm s}  \equiv \beta^{1/6} \lvert t \rvert^{-1} \simeq \beta^{1/6}\bar{\phi}(t)\,, 
\ee
which we identify as the strong coupling scale.  We now study the limit $\beta \rightarrow \infty$, $\lvert t\rvert\rightarrow \infty$, keeping $\Lambda_{\rm s}$ fixed.  Only the ordinary galileon terms \cite{Nicolis:2008in} survive, with spatial gradients suppressed by $\gamma$, so we scale them in taking the limit so that the limiting theory looks Lorentz invariant.  Because of the galileon non-renormalization theorem ~\cite{Luty:2003vm,Nicolis:2004qq,Hinterbichler:2010xn}, it follows that if we work at finite $\beta$, radiative corrections to $C_1,\ldots,C_5$ must be suppressed by powers of $1/\beta$, so the hierarchy we have set up is stable. 
Loop corrections also produce higher-derivative terms suppressed by $\Lambda_{\rm s}$, but these are consistently small at low energy so we have a derivative expansion in $\partial/\Lambda_{\rm s}$.

Finally, we discuss the issue of superluminality around the Poincar\'e-invariant vacuum $\phi = \bar{\phi}_0$. With $C_3\neq 0$, weak deformations
of this background exhibit superluminal propagation~\cite{Nicolis:2009qm}. (Our conditions cannot be simultaneously
satisfied with $C_3 = 0$.) Following the arguments of~\cite{Nicolis:2009qm}, superluminal effects can be consistently ignored
in the effective theory if the cutoff is sufficiently low: $\Lambda_0 \;\lsim\; \bar{\phi}_0/\sqrt{|C_3|} \sim \bar{\phi}_0/\sqrt{\beta}$. By relativistic and conformal invariance, the cutoff around
any background scales as $\Lambda \sim \phi/\gamma$. For consistency of our analysis, the lowest allowed cutoff around the NEC-violating solution is set by the mass of $\varphi$, namely
$1/|t|$. This implies $\Lambda_0 \sim \beta \bar{\phi}_0$, hence superluminal effects lie within the effective theory. 

In this paper we have shown that the NEC can be violated in a stable manner with subluminal perturbations,
from a theory which also admits a stable, Poincar\'e invariant vacuum. This represents a marked improvement
over earlier attempts, though the issue of superluminality around deformations of Poincar\'e remains~\cite{footnote2}.

{\em Acknowledgments}:  We thank P.~Creminelli, L.~Hui, A.~Nicolis, E.~Trincherini and M.~Trodden for helpful discussions. This work is supported in part by
the US DOE (G.E.J.M.), NASA ATP grant NNX11AI95G (A.J.), and NSF CAREER Award PHY-1145525 (J.K.).
Research at Perimeter Institute is supported by the Government of Canada through Industry Canada and by the Province of Ontario through the Ministry of Economic Development and Innovation.  This work was made possible in part through the support of a grant from the John Templeton Foundation. The opinions expressed in this publication are those of the authors and do not necessarily reflect the views of the John Templeton Foundation (K.H.).  



\begin{thebibliography}{99}

\bibitem{Gasperini:1992em} 
  M.~Gasperini and G.~Veneziano,
  Astropart.\ Phys.\  {\bf 1}, 317 (1993)
    [\href{http://arxiv.org/abs/hep-th/9211021}{hep-th/9211021}].

\bibitem{Khoury:2001wf} 
  J.~Khoury, B.~A.~Ovrut, P.~J.~Steinhardt and N.~Turok,
  Phys.\ Rev.\ D {\bf 64}, 123522 (2001)
  [\href{http://arxiv.org/abs/hep-th/0103239}{hep-th/0103239}].

\bibitem{Rubakov:2009np} 
  V.~A.~Rubakov,
  JCAP {\bf 0909}, 030 (2009).

\bibitem{Hinterbichler:2011qk} 
  K.~Hinterbichler and J.~Khoury,
  JCAP {\bf 1204}, 023 (2012).

\bibitem{Hinterbichler:2012mv} 
  K.~Hinterbichler, A.~Joyce and J.~Khoury,
  JCAP {\bf 1206}, 043 (2012)
  [\href{http://arxiv.org/abs/1202.6056}{arXiv:1202.6056 [hep-th]}].
 
\bibitem{Nayeri:2005ck} 
  A.~Nayeri, R.~H.~Brandenberger and C.~Vafa,
  Phys.\ Rev.\ Lett.\  {\bf 97}, 021302 (2006)
  [\href{http://arxiv.org/abs/hep-th/0511140}{hep-th/0511140}].
 
\bibitem{Creminelli:2010ba} 
  P.~Creminelli, A.~Nicolis and E.~Trincherini,
  JCAP {\bf 1011}, 021 (2010)
  [\href{http://arxiv.org/abs/1007.0027}{arXiv:1007.0027 [hep-th]}].
  
\bibitem{Dubovsky:2005xd}
  S.~Dubovsky, T.~Gregoire, A.~Nicolis and R.~Rattazzi,
  JHEP {\bf 0603} (2006) 025
  [\href{http://arxiv.org/abs/hep-th/0512260}{hep-th/0512260}].

\bibitem{ArkaniHamed:2003uy} 
  N.~Arkani-Hamed, H.~-C.~Cheng, M.~A.~Luty and S.~Mukohyama,
  JHEP {\bf 0405}, 074 (2004).

\bibitem{Creminelli:2006xe} 
  P.~Creminelli, M.~A.~Luty, A.~Nicolis and L.~Senatore,
  JHEP {\bf 0612}, 080 (2006)
  [\href{http://arxiv.org/abs/hep-th/0606090}{hep-th/0606090}].

\bibitem{Buchbinder:2007ad} 
  E.~I.~Buchbinder, J.~Khoury and B.~A.~Ovrut,
  Phys.\ Rev.\ D {\bf 76}, 123503 (2007)
  [\href{http://arxiv.org/abs/hep-th/0702154}{hep-th/0702154}].

\bibitem{Creminelli:2007aq} 
  P.~Creminelli and L.~Senatore,
  JCAP {\bf 0711}, 010 (2007).
  
\bibitem{Nicolis:2008in} 
  A.~Nicolis, R.~Rattazzi and E.~Trincherini,
  Phys.\ Rev.\ D {\bf 79}, 064036 (2009)
  [\href{http://arxiv.org/abs/0811.2197}{arXiv:0811.2197 [hep-th]}].
  
\bibitem{Creminelli:2012my} 
  P.~Creminelli, K.~Hinterbichler, J.~Khoury, A.~Nicolis and E.~Trincherini,
  \href{http://arxiv.org/abs/1209.3768}{arXiv:1209.3768 [hep-th]}.
    
\bibitem{deRham:2010eu} 
  C.~de Rham and A.~J.~Tolley,
  JCAP {\bf 1005}, 015 (2010).
  
\bibitem{Goon:2011qf} 
  G.~Goon, K.~Hinterbichler and M.~Trodden,
  JCAP {\bf 1107}, 017 (2011)
  [\href{http://arxiv.org/abs/1103.5745}{arXiv:1103.5745 [hep-th]}].
  
\bibitem{Dubovsky:2006vk}
  S.~L.~Dubovsky and S.~M.~Sibiryakov,
  Phys.\ Lett.\ B {\bf 638} (2006) 509
  [\href{http://arxiv.org/abs/hep-th/0603158}{hep-th/0603158}].

\bibitem{yizen}
L.~Berezhiani, Y.-Z.~Chu, J.~Khoury and M.~Trodden, in progress.

\bibitem{footnote1}
The brane tension $c_1$ will turn out to be positive for our choice of parameters.

\bibitem{nimaUVIR}
  A.~Adams, N.~Arkani-Hamed, S.~Dubovsky, A.~Nicolis and R.~Rattazzi,
  JHEP {\bf 0610}, 014 (2006).

\bibitem{Komargodski:2011vj} 
  Z.~Komargodski and A.~Schwimmer,
  JHEP {\bf 1112}, 099 (2011).
 [\href{http://arxiv.org/abs/1107.3987}{arXiv:1107.3987 [hep-th]}].

\bibitem{Martin:1965jj} 
  A.~Martin,
  Nuovo Cim.\ A {\bf 42}, 930 (1965).
  
\bibitem{Nicolis:2009qm} 
  A.~Nicolis, R.~Rattazzi and E.~Trincherini,
  JHEP {\bf 1005}, 095 (2010)
  [Erratum-ibid.\  {\bf 1111}, 128 (2011)].


\bibitem{Hinterbichler:2012fr} 
  K.~Hinterbichler, A.~Joyce, J.~Khoury and G.~E.~J.~Miller,
  \href{http://arxiv.org/abs/1209.5742}{arXiv:1209.5742 [hep-th]}.

\bibitem{Khoury:2012dn} 
  J.~Khoury, B.~A.~Ovrut and J.~Stokes,
  JHEP {\bf 1208}, 015 (2012)
  [\href{http://arxiv.org/abs/1203.4562}{arXiv:1203.4562 [hep-th]}].
  
\bibitem{Luty:2003vm} 
  M.~A.~Luty, M.~Porrati and R.~Rattazzi,
  JHEP {\bf 0309}, 029 (2003)
 \href{http://arxiv.org/abs/hep-th/0303116}{[hep-th/0303116]}.
  
\bibitem{Nicolis:2004qq} 
  A.~Nicolis and R.~Rattazzi,
  JHEP {\bf 0406}, 059 (2004).

\bibitem{Hinterbichler:2010xn} 
  K.~Hinterbichler, M.~Trodden and D.~Wesley,
  Phys.\ Rev.\ D {\bf 82}, 124018 (2010).
 [\href{http://arxiv.org/abs/1008.1305}{arXiv:1008.1305 [hep-th]}].

\bibitem{footnote2}
This may be avoided by considering more general bulk geometries which preserve scale invariance but break
special conformal transformations.

\end{thebibliography}
\end{document}